\RequirePackage{fix-cm}
\documentclass[smallextended]{svjour3}
\smartqed  
\usepackage{mathptmx}      

\usepackage{graphicx}
\usepackage{cite}
\usepackage{amsmath}
\usepackage{natbib}
\usepackage{color}
\usepackage{soul}
\usepackage{breakcites}
\usepackage{lineno}

\newcommand{\rev}[3]{#3}
\definecolor{grn}{rgb}{0,0.45,0}

\usepackage[normalem]{ulem}

\newcommand\canbret{\bgroup\markoverwith{\textcolor{grn}{\rule[0.5ex]{2pt}{0.7pt}}}\ULon}

\journalname{Journal of Mathematical Biology}

\begin{document}

\title{A Note on Species Richness and the Variance of Epidemic Severity}
\thanks{The authors thank NSF for partial support (Award Number: 1316334), as well as Profs. Greg Dwyer and Joe Mihaljevic for helpful discussions.}

\author{Peter Shaffery$^{1,\ast}$\\
Bret D. Elderd$^{2}$\\
Vanja Dukic$^{1}$}

\institute{Peter Shaffery \at
  University of Colorado Boulder, Dept of Applied Math\\
  Engineering Center, ECOT 225\\
  526 UCB Boulder, CO 80309-0526\\
  Tel.: +978-394-1443\\
  \email{pesh5067@colorado.edu}
  \and
  Bret Elderd \at
  Louisiana State University
  \and
  Vanja Dukic \at
  University of Colorado Boulder
}

\date{Received: date / Accepted: date}
  
\maketitle

\begin{abstract}
The commonly observed negative correlation between the number of species in an ecological community and  disease risk, typically referred to as ``the dilution effect'', has received a substantial amount of attention over the past decade. Attempts to test this relationship experimentally have revealed that, in addition to the mean disease risk decreasing with species number, so too does the \textit{variance} of disease risk. This is referred to as the ``variance reduction effect'', and has received relatively little attention in the disease-diversity literature.  Here, we set out to clarify and quantify some of these relationships in an idealized model of a randomly assembled multi-species community undergoing an epidemic. We specifically investigate the variance of the community disease reproductive ratio, a multi-species extension of the basic reproductive ratio $R_0$, for a family of random-parameter community SIR models, and   show how the variance of community $R_0$ varies depending on whether transmission is density or frequency-dependent. We finally  outline areas of further research on how changes in variance affect transmission dynamics in other systems.
\subclass{92 \and 15 \and 37}
\end{abstract}

\section{Introduction}
\label{intro}
A major focus of research efforts in the field of disease ecology is the effect of  biodiversity on epidemic behavior or pathogen prevalence (``disease-diversity" relationships). The most commonly discussed such relationship is the ``dilution effect'', which is an observed negative correlation between the number of different host species (species richness) and the incidence or risk of a specific disease \citep{ostfeld_biodiversity_2000-1,civitello_biodiversity_2015}.  The idea that greater biodiversity inhibits disease risk is attractive, intuitive, and easily reinforced by occasional catastrophic outbreaks in monocultures \citep{curl_control_1963,vandermeer_ecology_1992}. However, evidence exists to suggest that opposite ``amplification effects'' can sometimes also occur, where increased species richness results in an increase in disease risk \citep{salkeld_meta-analysis_2013,randolph_pangloss_2012}.  

\rev{2}{1}{While dilution and amplification have received the bulk of attention, higher-order relationships between disease incidence have recently started to become of interest  as well \citep{buhnerkempe_eight_2015,johnson_frontiers_2015}. One such property, the focus of this work, is the ``variance reduction effect", which refers to a phenomenon in which the variability of disease risk (as measured by either pathogen prevalence or the community disease reproductive ratio) decreases as species richness increases. 
Understanding this and other higher-order phenomena is important as the focus on mean behavior in disease ecology may not only miss important diversity relationships, but also result in biased inference and predictions. Within a single population, susceptibility to the pathogen or infectivity often varies a great deal amongst individuals \citep{dwyer_host_1997, lloyd-smith_superspreading_2005}.  When considering susceptibility, individual variability leads to a decidedly non-linear relationship in infection dynamics such that as pathogen levels increase transmission asymptotes at ever higher pathogen amounts \citep{elderd_population-level_2013}. Standard models that do not account for variation in susceptibility show a linear relationship to an increase in pathogen, which does not always hold when confronted with the empirical data \citep{dwyer_host_1997}. At the between-population scale, variance reduction suggests that diverse host communities may be less susceptible to extreme epidemics, and that their dynamics are more predictable and perhaps controllable. In addition, understanding when to expect variance reduction, or if other variance effects are possible or even likely at play,
can help to determine the role that diversity plays in epidemic modeling and inference as well as how variance reduction effects may help stabilize communities.}

While the variance reduction effect has been observed in randomized experiments, so far the underlying mechanism is not precisely understood \citep{mitchell_effects_2002, rottstock_higher_2014, mihaljevic_scaling_2014}.  The most popular explanation argues that the variance reduction effect is a ``selection effect'' \citep{huston_hidden_1997}; that communities with increased numbers of species are more likely to have a high degree of overlap, and therefore share properties which result in a similar response to disease. This results in a decrease in the variability of associated disease risk in species rich communities as compared to species poor communities.

Nevertheless, controlling for the effect of community composition (eg. average pathogen susceptibility of the member host species) in experimental communities does not appear to remove the variance reduction effect, as would be expected if a selection effect were the driving mechanism \citep{mitchell_effects_2002}.  This suggests that the selection effect explanation of variance reduction is incomplete and that the reduced variance in pathogen prevalence among species rich communities is due to some other factors besides increased similarity among host populations. Furthermore, variance reduction also appears \textit{in silico}, in simulations of dynamical models of epidemic spread with random parameters \citep{mihaljevic_scaling_2014}. As these models allow for explicit control of all variables, it is therefore unlikely that the unexplained variance reduction is due to a confounding variable, as was proposed in \citet{mitchell_effects_2002}.

In this work we set out to explore the mathematical properties of variance reduction using a community, Susceptible-Infected-Recovered (SIR) model
of an epidemic which infects a community of randomly selected species. For this model we derive and analyze the ``Next Generation Matrix'' (NGM) and its spectral norm, which we term the ``community invasion factor''. This definition of community $R_0$ approximates the average rate of new infections per individual, generalizing the reproductive ratio from classical epidemic modeling \citep{diekmann_definition_1990}.

 Due to the randomization of community host species, our NGM is a random  matrix and so its spectral norm will also be a random variable. We derive bounds for the variance of this distribution and thus the community $R_0$, which depend on the species richness of the host community. In doing so, we show that the presence and strength of a variance reduction phenomenon hinges on the pathogen transmission mode (density or frequency-dependent), and that when disease is density dependent a variance amplification (an increase in variance with species richness) is also possible.

\section{Model Formulation}
\label{sec:mod_form}
\subsection{Epidemic Model}
\label{subsec:epi_mod}
We consider a random-parameter community epidemic model, with SIR dynamics describing the course of the disease within and between $n$ host species. Each species within the community is undergoing an epidemic of the same pathogen, according to the SIR model.  Our choice of model and assumptions follows \citet{dobson_population_2004}, and reflects a balance of realism and tractability. 

Our multi-host SIR epidemic model is:
\begin{equation}
\begin{split}
\dot{S}_i &= b_i (S_i + I_i + D_i) - p_i \sum\limits_{j=1}^n \beta_{ij} I_j - d_i S_i\\
\dot{I}_i &= p_i \sum\limits_{j=1}^n \beta_{ij} I_j - (d_i + \alpha_i) I_i\\
\dot{D}_i &= \alpha_i I_i - d_i D_i
\label{eq:model}
\end{split}
\end{equation}
and it assumes that for host species $i$, living individuals fall into one of three categories: susceptible ($S_i$), infected ($I_i$), or recovered ($D_i$), and that individuals of all categories reproduce at rate $b_i$ and die at rate $d_i$.  For convenience, however, we assume that $b_i = d_i$. 

For the between-species transmission rates, we use the convention that $\beta_{ij} = c_{ij} \frac{\beta_{ii} + \beta_{jj}}{2}$ where $c_{ij} \geq 0$. Note that we let $c_{ij} \neq c_{ji}$, which allows for transmission to be asymmetric between species. The $p_i$ factors reflect whether transmission is density-dependent ($p_i=S_i$) or frequency-dependent ($p_i=\frac{S_i}{\sum_i N_i}$). Density dependence refers to transmission dynamics where the rate of infectious contacts increases in proportion to population size, whereas frequency dependence assumes that it is constant. A canonical example of the former would be the flu, and of the latter would be a sexually transmitted infection. See Table 1 for the definitions of all model parameters in the model Eq. \ref{eq:model}.

To account for the randomization employed in experiments, we treat the parameters of the ODEs in (\ref{eq:model}) as random variables, \rev{AE}{2}{with the exception of the $c_{ij}$, which we assume to be fixed givens. Reserving the $c_{ij}$ as fixed allows us to account for different regimes of between-species transmission explicitly. In our proof and simulation we will assume that all $c_{ij} = c$, and by varying $c$ we can look at variance reduction in communities with overall low- and high- between-species infection. Furthermore, allowing for large numbers of the $c_{ij}$ to be set equal to $0$ (as would be the case with a vector-borne pathogens) may constrain variance reduction more than the bound derived below, although we do not deal with this possibility here.}

Based on this random ODE community SIR model, we will derive bounds for the variance of disease risk as a function of species richness, under different assumptions about the transmission mode of the disease.

\begin{table}
\begin{tabular}{lll}
\hline\noalign{\smallskip}
Parameter & Definition\\
\noalign{\smallskip}\hline\noalign{\smallskip}
$b_i$ & Per capita birth rate of host species $i$\\
$d_i$ & Per capita death rate of host species $i$\\
$\beta_{ij}$ & Per capita rate of pathogen transmission from host species $i$ to host species $j$\\
$\alpha_i$ & Recovery rate of host species $i$ when infected with the pathogen\\
$K_i$ & \rev{1}{1.1.1}{Population size of species $i$ in the absence of the pathogen}\\
$N_i$ & \rev{1}{1.1.2}{Total population size ($N_i=S_i+I_i+R_i$)}\\
$c_{ij}$ & Parameter that modifies between-species transmission from an arithmetic mean of their within-species rates\\
\hline
\noalign{\smallskip}\hline
\end{tabular}
\caption{Definitions of parameters used in the model in equation~\ref{eq:model}.}
\end{table}

\subsection{Measures of  Epidemic Severity}
\label{subsec:epi_sev}
\subsubsection{\rev{1}{1.2.1}{The Next Generation Matrix}}
\label{subsubsec:ngm}
For many classes of multi-host epidemic models the severity of the epidemic can be captured through the  ``Next Generation Matrix" (NGM) \citep{diekmann_definition_1990}.  Denoting  the NGM matrix as  $\pmb{G}$, its elements $G_{ij}$ can be interpreted as the expected number of new infections an individual of type $j$ produces among susceptibles of type $i$, over its entire infectious duration. Letting $\phi_i^{(1)}$ denote the number of infected individuals of type $i$ in the first generation during the initial disease outbreak, then the expected number of infected individuals of type $i$ in the second generation is $\phi_i^{(2)} = \sum_j G_{ij} \phi_j^{(1)}$.
For the entire community, we have that $\pmb{G}$ pushes the initial outbreak vector forward to the second generation vector, $\pmb{\phi}^{(2)} = \pmb{G} \pmb{\phi}^{(1)}$, where $\pmb{\phi}^{(m)} = [\phi_1^{(m)},...,\phi_n^{(m)}]^T$.

For our model, the NGM can be derived from the Jacobian $\pmb{J}$ of the $\dot{I}_i$ with respect to all $I_j$, evaluated at $I_j = 0$, the so-called disease-free equilibrium. Letting $s(\pmb{J}) = \text{max}(\{\text{Re}(\lambda), \lambda \in \sigma(\pmb{J})\})$ where $\sigma(\pmb{J})$ is the spectrum of $\pmb{J}$, the disease-free equilibrium is unstable whenever $s(\pmb{J}) > 0$.  When the disease-free equilibrium is unstable then small outbreaks of the disease will spread into full epidemics.

For the system in Eqs. (\ref{eq:model}), direct calculation gives that $J_{ij} = \beta_{ij} p_i - (d_i + \alpha_i) \delta(i-j)$, where $\delta(i-j)$ is the Kronecker delta function.  This implies that $\pmb{J} = \pmb{T} - \pmb{\Sigma}$, where $\pmb{T}$ is referred to as the ``transmission matrix'' and represents the rate at which new infections are being created. $\pmb{\Sigma}$ is a diagonal matrix whose nonzero elements are $(d_i + \alpha_i)$ and is referred to as the ``recovery matrix'', and represents that rate at which infected individuals recover.  Given this partition of $\pmb{J}$ we define the NGM as $\pmb{G} = \pmb{T} \pmb{\Sigma}^{-1}$; intuitively this can be thought of as ``rate of infection over rate of recovery".  

It can be quickly verified that the elements of $\pmb{G}$ for Eqs. (\ref{eq:model}) are $G_{ij} = \frac{\beta_{ij}}{d_i + \alpha_i}$.  However, given our assumption that $\beta_{ij} = \frac{c_{ij}}{2}(\beta_{ii} + \beta_{jj})$ it is convenient to reparameterize $\pmb{G}$ in terms of its diagonal elements $R_i = \frac{\beta_{ii}}{\theta_i}$, where $\theta_i = d_i + \alpha_i$. The NGM for a community with $n$ host species then has the form:
\begin{equation}
\pmb{G}_n = \left[
\begin{array}{ccc}
p_1 R_1 & \dots & \frac{c_{1n} p_1}{2} \left( R_1 + \frac{\theta_n}{\theta_1} R_n \right)  \\
\vdots & \ddots & \vdots\\
\frac{c_{n1} p_n}{2} \left( R_n + \frac{\theta_1}{\theta_n} R_1 \right) & \dots & p_n R_n \\
\end{array}
\right]
\label{eq:NGM}
\end{equation}
\rev{1}{1.1.2}{Observe that, at the disease-free equilibrium, $p_i = K_i$ when transmission is density-dependent and $p_i = \frac{K_i}{\sum_i K_i}$ when transmission is frequency-dependent.}

\subsubsection{\rev{1}{1.2.1}{Community $R_0$ and Community Invasion Factor}}
\label{subsubsec:comm_r}
Having calculated the NGM $\pmb{G}_n$, the question now is how best to extract information from it about the severity of the epidemic. The canonical choice would be to take its spectral radius, $\rho(\pmb{G}_n)$, as was done by Diekmann et. al. in their 1990 paper on the definition of $R_0$ for heterogeneous populations. This choice was motivated by the observation that, for any matrix norm $|| \cdot ||$, $\rho(\pmb{G}_n) = \lim_{m \rightarrow \infty} ||\pmb{G}_n^m||^{\frac{1}{m}}$, so in a certain sense the spectral radius of $\pmb{G}_n$ captures the average, long-term growth factor of the initial outbreak.  Furthermore, it can be shown that $\text{sign}(s(\pmb{J})) = \text{sign}( \rho(\pmb{G}_n) - 1)$, where $s(\pmb{J})$ denotes the modulus of the largest eigenvalue of $J$. This means that the spectral radius of $\pmb{G}_n$ also captures the stability of the disease-free equilibrium. For brevity we do not include a proof of this fact, but a straightforward one can be found in the appendix of \citet{diekmann_construction_2009}.

 In its original derivation, however, the NGM is derived as an infinite-dimensional forward operator $\mathcal{G}_n$ for the dynamical system underlying the epidemic model \citep{diekmann_definition_1990}. It's repeated application, denoted $\mathcal{G}_n^m$, produces the exact trajectory of the epidemic for discrete time-steps $m$. However in \ref{subsubsec:ngm} our derivation of $\pmb{G}_n$ is based on a linearized version of the underlying model, and its iterated application does not produce the true epidemic, only a short-time approximation of it. From that perspective, the large-m behavior of $||\pmb{G}_n^m||^{\frac{1}{m}}$ is less relevant than when $m=1$.
 
Furthermore, the spectral radius $\rho(\pmb{G}_n)$ is an inconvenient definition of epidemic severity, for our purposes. The spectral radius lacks a number of properties found in matrix norms, which make it challenging to derive results about them for general random matrices. For example: spectral radii do not obey a triangle inequality; it is possible that $\rho(\pmb{M}+\pmb{N}) > \rho(\pmb{M}) + \rho(\pmb{N})$ for matrices $\pmb{M}$ and $\pmb{N}$. As we show, properties like a triangle inequality will be important in proving our results. 

For these reasons we will instead rely on the spectral norm $||G_n||_2$, to measure epidemic severity. This can be understood as measuring the maximum second generation growth factor of the infective population, since $||G_n||_2 = \max_{\phi} \frac{||\pmb{G} \pmb{\phi}^{(1)}||}{||\pmb{\phi}^{(1)}||}$. \rev{2}{2}{We therefore term this measure the ``community invasion factor", to distinguish it from the community $R_0$, $\rho(G_n)$.}

\subsubsection{Comparing Spectral Norm and Spectral Radius for Epidemic Random Matrices}
\label{subsubsec:spec_norm_rad}
We would like to assess whether our definition of epidemic severity $||\pmb{G}_n||$ is ``equivalent" to the canonical choice ($\rho(\pmb{G}_n)$). Equivalency of metrics can have several meanings, but here we will consider just two. First, we test the extent to which exchanging $\rho(\pmb{G}_n)$ and $||\pmb{G}_n||_2$ will preserve the ordering; i.e., whether for two random matrices $\pmb{G}_n$ and $\pmb{G}_n'$ with
$\rho(\pmb{G}_n) \leq \rho(\pmb{G}_n')$ we also have that $||\pmb{G}_n||_2 \leq ||\pmb{G}_n'||_2$ (termed ``monotonicity"). Second, we compare how effectively $\rho(\pmb{G}_n)$ and $||\pmb{G}_n||_2$ predict total epidemic size for numerically solved trajectories of epidemic model (\ref{eq:model}) (termed ``predictiveness").  If $\rho(\pmb{G}_n)$ and $||\pmb{G}_n||_2$ are equivalent in these ways, then we can meaningfully refer to ``high" or ``low" community $R_0$, regardless of which definition we use. 

To assess monotonicity and predictiveness we will look at the sampling distributions of $\rho(\pmb{G}_n)$, $||\pmb{G}_n||_2$, and numerical epidemic size under four cases: assuming either a small community ($n=5$) or a large community ($n=50$), and either realistic or unrealistic distributions for the $R_i$, $K_i$, and $\theta_i$. The realistic parameter distributions are drawn from natural systems and follow \citet{mihaljevic_scaling_2014}, whereas the unrealistic case assigned each parameter a Gamma density with randomly selected shape and scale parameters, uniformly chosen from the interval $[1,5]$. We set $c_{ij} = .05$ (following the value chosen in \cite{mihaljevic_scaling_2014}) and assume frequency dependence. \rev{1}{1.1.3}{Furthermore, we will assume that the birth and death rate is negligible over the duration of the epidemic, and set all $b_i=d_i=0$.}

\begin{figure}
  \includegraphics[scale=.3]{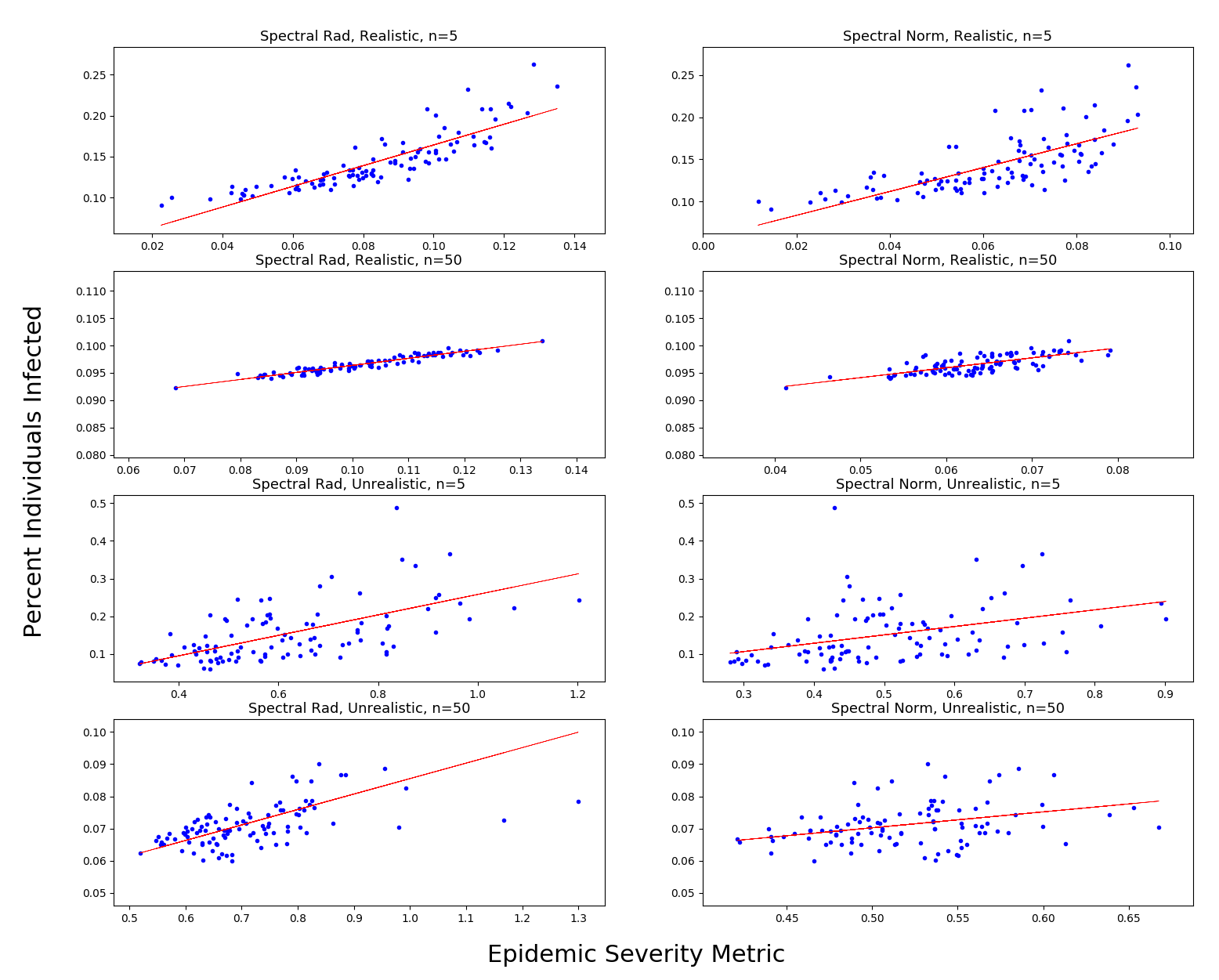}
  \caption{\small{Estimated coefficient of determination between percent infected individuals and $\rho(\pmb{G}_n)$ (spectral radius) and $||\pmb{G}_n||_2$ (spectral norm). Percent infected was calculated with \rev{AE}{1}{$M=100$} random draws of $\rho(\pmb{G}_n)$, $||\pmb{G}_n||_2$, and percent infected. Percent infected was calculated by numerically solving Eq. (\ref{eq:model}) using Runge-Kutta (4,5) for 100 time steps of size $\delta t = 1$, and with $b_i=d_i=0$. Subplot headers indicate epidemic metric (canonical ``spectral radius'' or proposed ``spectral norm''), underlying parameter distribution (realistic or unrealistic) as well as coefficient of determination ($R^2$). Linear fits and $R^2$ were calculated after excluding high-leverage points and outliers, which are plotted with an X.}}
  \label{fig:rad_norm_size}
\end{figure}

After drawing from these sampling distributions we quantify predictiveness using simple linear regression: we estimate the coefficient of determination, $R^2$ between $\rho(\pmb{G}_n)$ and total percent infected, and between $||\pmb{G}_n||_2$ and total percent infected (Fig. \ref{fig:rad_norm_size}). To prevent bias from highly influential points, these calculations exclude high-leverage points, those points $i$ with leverage values $h_i > \frac{2(m+1)}{n}$ (where $m=1$ is the number of independent variables in the model), as well as outliers, points with studentized residuals $\hat{\epsilon}_i = \frac{\epsilon_i}{\sigma_{(-i)} \sqrt{1-h_i}}$ outside the range [-3,3] (where $\epsilon_i = \hat{y}_i - y$ is the un-studentized residual and $\sigma_{(-i)}$ is the standard deviation of that observations $y_1,...,y_n$ excluding $y_i$). These points are still included in Fig. \ref{fig:rad_norm_size}, but are indicated with an X to show that they do not influence the linear fits.

\begin{figure}
  \includegraphics[scale=.25]{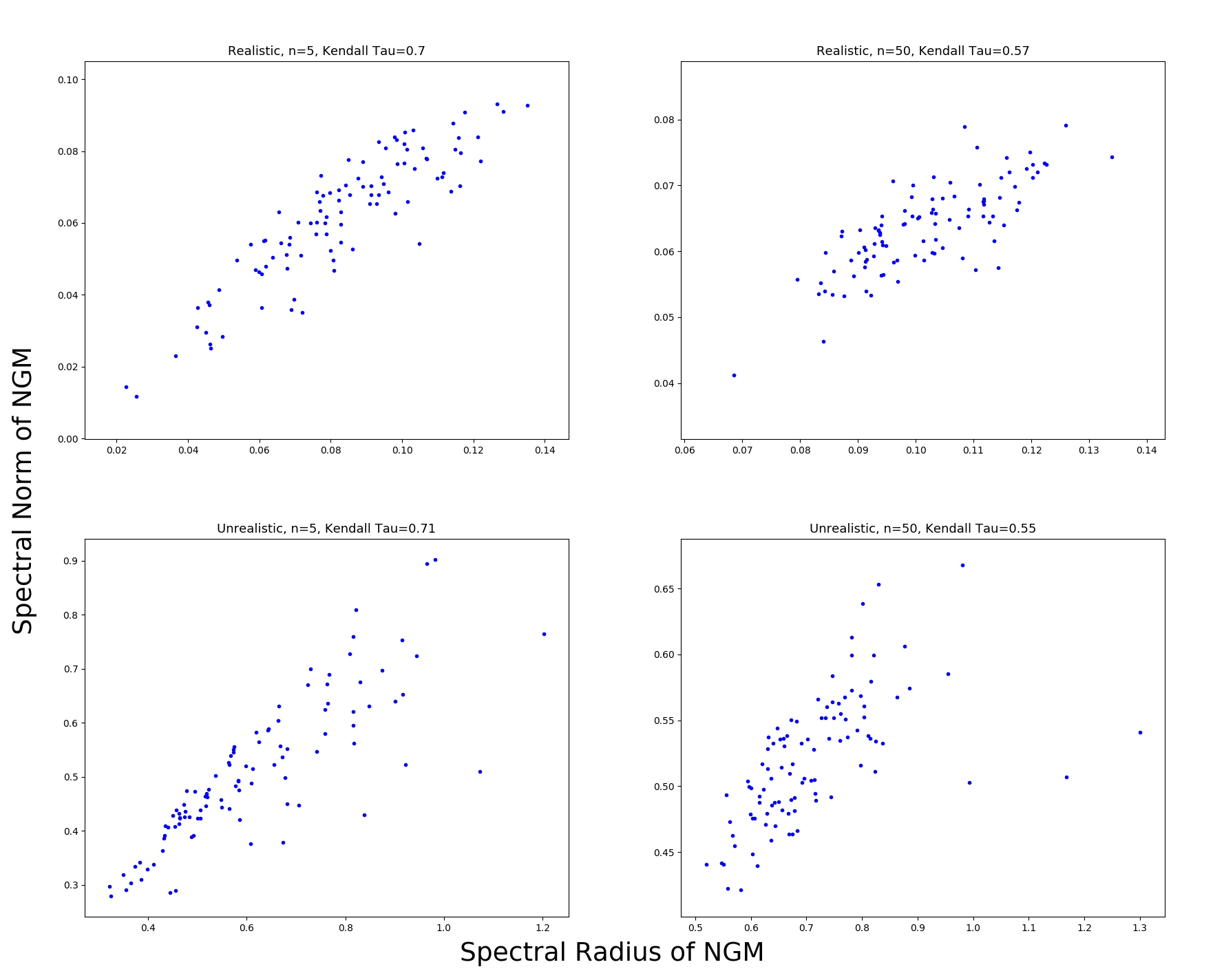}
  \caption{Estimated Kendall Tau and joint sampling distribution for $M=100$ samples of $\rho(\pmb{G}_n)$ and $||\pmb{G}_n||_2$. Subplot headers indicate underlying parameter distribution (realistic or unrealistic) as well as coefficient of determination ($R^2$).}
  \label{fig:kendall_tau}
\end{figure}

To quantify monotonicity we use the Kendall Tau rank correlation coefficient (Fig. \ref{fig:kendall_tau}). Kendall Tau tests the extent to which two random variables have a monotonic relationship, without specifying the functional form of that relationship. It is calculated by comparing pairs $(\rho(\pmb{G}_n),||\pmb{G}_n||_2)$ and $(\rho(\pmb{G}_n'),||\pmb{G}_n'||_2)$. If $\rho(\pmb{G}_n) \geq \rho(\pmb{G}_n')$ and $||\pmb{G}_n||_2 \geq ||\pmb{G}_n'||_2$ the comparison is said to be ``cordant", otherwise they are termed ``discordant". All pairs of samples are compared (for $n$ samples we make $n(n-1)/2$ comparisons) and the Kendall Tau coefficient is then calculated by $\tau := \frac{n_c - n_d}{n_c + n_d}$, where $n_c$ is the number of cordant comparisons and $n_d$ is the number of discordant ones. A value of $\tau = \pm 1$ implies a perfectly monotonic relationship (either increasing or decreasing, respectively).

From Fig. \ref{fig:rad_norm_size} we see that, for the realistic parameter distributions, both metrics predict epidemic severity fairly well, however the canonical choice $\rho(\pmb{G}_n)$ outperforms our candidate somewhat, for all cases. We attribute this to the stability interpretation of $\rho(G_n)$, which provides some information about the long-time behavior of the epidemic, whereas $||G_n||$ provides information purely regarding the short term growth rate. Fig. \ref{fig:kendall_tau} indicates that the relationship between $\rho(\pmb{G}_n)$ and $||\pmb{G}_n||_2$ is fairly monotonic (and increasing), as desired. \rev{AE}{3}{We note that Kendall Tau does decrease somewhat as species richness $n$ grows large, we understand this behavior by observing that when $n=1$ the spectral radius and spectral norm are exactly equal and so the Kendall Tau is 1. As $n$ grows, the increasing volume of possible matrices $G_n$ means that it becomes more probable for spectral norm to decrease while the radius increases, or vice versa.}

While these results do not indicate that our proposed metric of epidemic severity should replace $\rho(\pmb{G}_n)$, they do suggest that it will suffice for our analysis of variance reduction. Clearly $||\pmb{G}_n||_2$ is not as predictive of epidemic size as $\rho(\pmb{G}_n)$, however the correlation is strong enough that variance reduction effects present in $||\pmb{G}_n||_2$ should still manifest in direct measurement of epidemic spread. Furthermore, given its relationship with $\rho(\pmb{G}_n)$ it may be possible to reformulate our results for the canonical metric as well, although we will not do so here.

\section{Results and Proof}
\label{sec:results}
Going forward we will use the notation $r(\pmb{G}_n) = ||\pmb{G}_n||_2$. Recall that our model represents an approximation of natural community assembly processes wherein a host community is populated with randomly selected species from a global pool of potential members \citep{mitchell_effects_2002}. In our model each host species is uniquely defined by the random triple of positive reals $(R_i, \theta_i, K_i)$. While we allow for dependence between the members of a single triple, we assume that any two distinct triples (ie. species) are independent. With this randomness, our NGM (\ref{eq:NGM}) is thus a random matrix, and its spectral norm is a random variable, whose distribution is induced by the distribution of $(R_i, \theta_i, K_i)$. We can now define a variance reduction effect as occurring if $\text{Var}_{R_i,\theta_i,K_i}[r(\pmb{G}_n)]$ is decreasing in $n$, the number of unique host species.

Our main result is the following:
\begin{theorem}
  Let $\pmb{R} = [R_1,...,R_n]^T$, $\pmb{\theta} = [\theta_1,...,\theta_n]^T$, and $\pmb{K} = [K_1,...,K_n]^T$ be random, real valued vectors with independent elements.  Assume that for all $n$, the support of $\pmb{R}$, $\pmb{\theta}$, and $\pmb{K}$ is such that $R_i \leq \Gamma$, $ 0 < \Theta_1 \leq \theta_i \leq \Theta_2$, and $K_i \leq \kappa$.  Let $\pmb{G}(\pmb{R},\pmb{\theta}, \pmb{K})$ be a matrix-valued function defined in Eq.~\ref{eq:NGM}, under frequency-dependent transmission, ie. $p_i = \frac{K_i}{\sum\limits_i K_i }$.  Then $\text{Var}[ || \pmb{G}( \pmb{R}, \pmb{\theta}, \pmb{K} ) ||_2 ] \sim \mathcal{O}( n^{-1/2} + 1 )$.
  \label{th:main}
\end{theorem}

Before we begin the proof of this result, it is constructive to look at why we might intuitively expect it to be true.  First, in the case of frequency-dependent transmission, the $p_i \sim \mathcal{O}(n^{-1})$.  We can then ``pull out'' a factor of $\frac{1}{n}$ from $\pmb{G}_n$ by writing $\pmb{G}_n = \frac{1}{n} \pmb{G}_n'$.  This implies that $\text{Var}[r(\pmb{G}_n)] = \frac{1}{n^2} \text{Var}[r(\pmb{G}'_n)]$.  Second, we observe that for many classes of random matrices $\pmb{M}$ one has that $\text{Var}[r(\pmb{M})] \sim o(n)$. Intuitively this can be understood as a concentration of measure result \citep{tao_topics_2012}. An $n \times n$ matrix is a high-dimensional object, even for moderate $n$, and matrix norms are continuous in the matrix elements. As $n$ grows, a large change in the norm (or radius) requires increasingly large numbers of elements to experience substantial, ``coordinated'' deviations.  Such deviations are improbable for sufficiently large $n$, and so the bulk of the probability mass concentrates near the mean.

If such a result were to hold for $\pmb{G}'$, then asymptotically we would see $\text{Var}[r(\pmb{G}_n)] \sim \mathcal{O}(n^{-1})$.  However these results typically hold only for specific classes of random matrices, such as Hermitian matrices with independent, mean-zero random entries \citep{tao_topics_2012} or sums of fixed matrices with random coefficients \citep{anderson_introduction_2010}.  The NGM in (\ref{eq:NGM}) is not Hermitian, and it has dependence between its entries; random matrices with correlated elements can behave idiosyncratically \citep{schenker_semicircle_2005}.  Our proof then, amounts to showing that these shortcomings are not sufficient enough to overwhelm the concentration of measure (ie. the variance reduction effect) entirely, although they do reduce its severity.

\subsection{Outline of Proof}
\label{subsec:proof_outline}
We begin by properly defining $\pmb{G}_n'$:
\begin{equation}
  \pmb{G}'_n = \frac{1}{\bar{K}} \left[
    \begin{array}{ccc}
      K_1 R_1 & \dots & \frac{c_{1n} K_1}{2} \left( R_1 + \frac{\theta_n}{\theta_1} R_n \right)  \\
      \vdots & \ddots & \vdots\\
      \frac{c_{n1} K_n}{2} \left( R_n + \frac{\theta_1}{\theta_n} R_1 \right) & \dots & K_n R_n \\
    \end{array}
  \right].
  \label{eq:gprime}
\end{equation}
Where we have defined $\bar{K} = \frac{1}{n} \sum\limits_{i=1}^n K_i$, which is nearly constant for large $n$.  Observe that $\pmb{G}_n'$ is effectively the ``density-dependent version'' of \ref{eq:NGM} and therefore the effect of switching from density to frequency dependence is to induce the scaling $\pmb{G}_n = \frac{1}{n} \pmb{G}_n'$, and subsequently $\text{Var}[ \rho(\pmb{G}_n) ] = \frac{1}{n^2} \text{Var}[\rho(\pmb{G}_n')]$. This $\frac{1}{n}$ scaling is what produces the variance reduction effect; we therefore expect that a variance reduction effect will occur only in the frequency-dependent case. Absent this scaling, ie. when disease is density dependent, a variance amplification should occur.

For convenience we will adopt the notation $r(\pmb{R},\pmb{\theta},\pmb{K}) = ||\pmb{G}_n'(\pmb{R},\pmb{\theta},\pmb{K})||_2$.
Next we observe that, by the Law of Total Variance:
\begin{equation}
  \text{Var}[ r(\pmb{R},\pmb{\theta},\pmb{K}) ] = E_{\pmb{\theta},\pmb{K}} \left[ \text{Var}_{\pmb{R}}[ r(\pmb{R},\pmb{\theta},\pmb{K}) | \pmb{\theta},\pmb{K} ] \right] + \text{Var}_{\pmb{\theta},\pmb{K}} \left[ E_{\pmb{R}}[ r(\pmb{R},\pmb{\theta},\pmb{K}) | \pmb{\theta},\pmb{K} ] \right]
  \label{eq:totvar}
\end{equation}

Our proof will show that the first term on the right hand side of Eq. \ref{eq:totvar} grows no faster than $\mathcal{O}(n^{3/2})$, while the second term is bounded by $\mathcal{O}(n^2)$. The desired result then follows from the $\frac{1}{n^2}$ scaling due to frequency dependence.

\subsection{Bounding $E_{\theta,K} \left[ \text{Var}_{R}[ r(R,\theta,K) | \theta,K ] \right]$ }
\label{subsec:term1_bound}
Our result follows from the following theorem:
\begin{theorem}[Concentration Inequality for Lipschitz Functions]
  Let $X_1,...,X_n$ be independent, real-valued random variables with $|X_i| \leq A$, for all $1 \leq i \leq n$.  Let $F: \mathbf{R}^n \rightarrow \mathbf{R}$ be a convex, Lipschitz function with Lipschitz constant $L$.  Then for any $t$ one has that:
  \[
    P[ ( F(X) - E[F(X)] )^2 \geq t ] \leq \exp(\frac{t}{2 A^2 L^2})
  \]
  \label{th:CILF}
\end{theorem}
\sloppy
This result is a well-known consequence of Talagrand for 1-Lipschitz functions \citep{boucheron_concentration_2013}, that has here been scaled to accommodate L-Lipschitz functions.  We observe that, since $\pmb{G}(\pmb{R}+\pmb{R}',\pmb{\theta},\pmb{K}) = G(\pmb{R},\pmb{\theta},\pmb{K}) + G(\pmb{R'},\pmb{\theta},\pmb{K})$,   then $r(\pmb{R},\pmb{\theta},\pmb{K})$ satisfies the norm axioms over $\pmb{R}$.  Since norms are equivalent over finite dimensions this immediately implies that $r(\pmb{R} - \pmb{R}',\pmb{\theta},\pmb{K}) \leq L_{\pmb{\theta},\pmb{K}} ||\pmb{R} - \pmb{R}'||_2$, ie. that $r(\pmb{R},\pmb{\theta},\pmb{K})$ is Lipschitz in $\pmb{R}$ with respect to $||\pmb{R}||_2$.  Simple integration of the right hand side of Thm \ref{th:CILF}, together with $R_i \leq \Gamma$, gives the following:
\[
  \begin{split}
    \text{Var}_{\pmb{R}}[ r(\pmb{R},\pmb{\theta},\pmb{K})|\pmb{\theta},\pmb{K}] \leq 2 \Gamma^2 L_{\pmb{\theta},\pmb{K}}^2 \\
  \end{split}
\]

A straightforward calculation shows that $L_{\pmb{\theta},\pmb{K}}^2$ is bounded by $C \sim \mathcal{O}(n^{3/2})$ (see Appendix \ref{subsec:app_bound}), where $C$ is independent of $\pmb{\theta}$ and $\pmb{K}$. We can therefore bound $ E_{\pmb{\theta},\pmb{K}} \left[ \text{Var}_{\pmb{R}}[ r(\pmb{R},\pmb{\theta},\pmb{K}) | \pmb{\theta},\pmb{K} ] \right]$ by the same term, which concludes the proof.

\subsection{Bounding $\text{Var}\left[ E[ r(R,\theta,K) | \theta,K ] \right]$ }
\label{subsec:term2_bound}
Here we will use the fact that $r(\pmb{R},\pmb{\theta},\pmb{K}) \leq ||\pmb{G}'(\pmb{R},\pmb{\theta},\pmb{K})||_F$, where $|| \pmb{M} ||_{F} = \sqrt{ \sum\limits_{i,j} M_{ij}^2}$ is the Frobenius matrix norm.  Therefore $Var[ E[r(\pmb{R},\pmb{\theta},\pmb{K}) | \pmb{\theta},\pmb{K}] ] \leq E[ E[r(\pmb{R},\pmb{\theta},\pmb{K}) | \pmb{\theta},\pmb{K}]^2 ] \leq E[ ||\pmb{G}'(\pmb{R},\pmb{\theta},\pmb{K})||_F^2 ]$. The last term on the right hand side is simply the expectation of a sum of bounded terms, and hence grows no faster than $\mathcal{O}(n^2)$.  We therefore have that $Var[ E[r(\pmb{R},\pmb{\theta},\pmb{K}) | \pmb{\theta},\pmb{K}] ] \sim \mathcal{O}(n^2)$, as desired.

\subsection{Numerical Experiments}
\label{subsec:num_exp}
To test the extent to which the above derived bounds are indeed useful for understanding of the actual behavior of $\text{Var}[||\pmb{G}(\pmb{R},\pmb{\theta},\pmb{K})||_2]$, we performed some simple Monte Carlo simulations of random realizations of the spectral norm of our NGM (\ref{eq:NGM}) with realistic distributions of parameter values for the case of frequency-dependent transmission, following \citet{mihaljevic_scaling_2014}.

We assumed that the joint distribution of $K_i,R_i,$ and $\theta_i$ was such that $R_i$ and $\theta_i$ were conditionally independent, given $K_i$. $K_i$ was then sampled from a log-normal distribution, in accordance with Preston's law of abundance distributions \citep{verberk_explaining_2011}.  Given $K_i$ we then sampled $\theta_i$ and $R_i$ from truncated normal distributions (with support over $[.1,5]$ and $[0,10]$, respectively), conditioned on the values $k_i = \ln(K_i)$. The mode of the $\theta_i$ was proportional to $\exp ( k_i - 2 )$, while for the $R_i$ it is simply proportional to $k_i$.  The constant of proportionality between $\theta_i$ and $\exp(k_i-2)$ is well established, following relationships between species abundance and life span.  \rev{1}{1.1.5}{In addition, we allow for correlations between species susceptibility $R_i$ and log-log-abundance $\ln(k_i)$, which we denote $a_R$. Although we did not explicitly encode ``pace-of-life" traits into this model, the parameter $a_R$ allows us to implictly consider their impact on variance reduction as fast-lived species often have high equilibrium abundances \citep{dobson_population_2004}} 

Using these sampling distributions we estimated separately each term in Eq. (\ref{eq:totvar}): the expectation of the conditional variance ($E_{\pmb{\theta},\pmb{K}} \left[ \text{Var}_{\pmb{R}}[ r(\pmb{R},\pmb{\theta},\pmb{K}) | \pmb{\theta},\pmb{K} ] \right]$) and the variance of the conditional expectation ($\text{Var}\left[ E[ r(\pmb{R},\pmb{\theta},\pmb{K}) | \pmb{\theta},\pmb{K} ] \right]$) for both frequency- and density-dependent NGMs.  Furthermore, for each type of NGM we considered four test cases: both a low and medium susceptibility-abundance correlation case ($a_R = .5$ and $a_R=2$). \rev{AE}{2}{, as well as a low and high between-species infection case using $c_{ij}=.05$, and $c_{ij}=1$. These are the values used in \cite{mihaljevic_scaling_2014} and the upper limit for the $c_{ij}$ discussed in \cite{dobson_population_2004}, respectively.}

\begin{figure}
  \includegraphics[scale=.6]{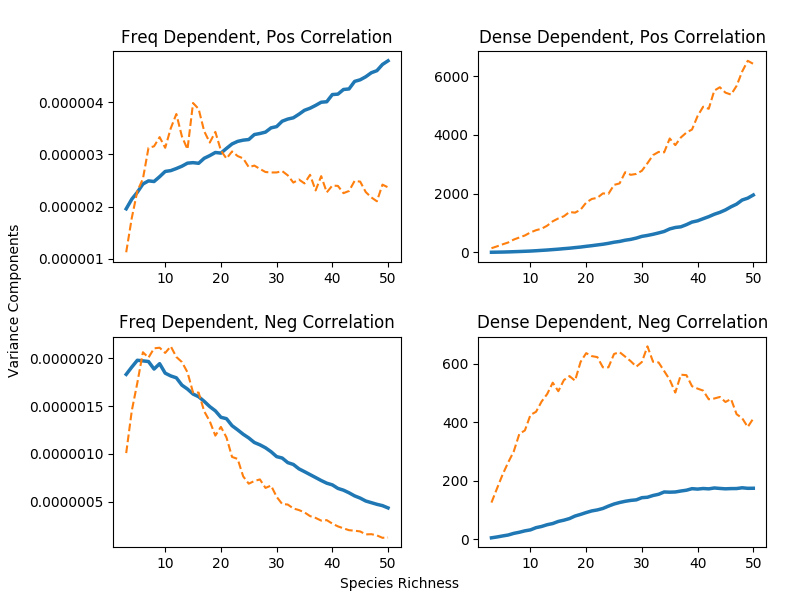}
  \caption{ Monte Carlo estimates of the mean conditional variance (blue, solid line) and the variance of the conditional mean (yellow, dashed line), the first and second terms (respectively) in Eq. (\ref{eq:totvar}), for the case where susceptibility exhibits positive or negative correlation with increases in species richness.}
  \label{fig:rev2_case}
\end{figure}

\begin{figure}
  \includegraphics[scale=.4]{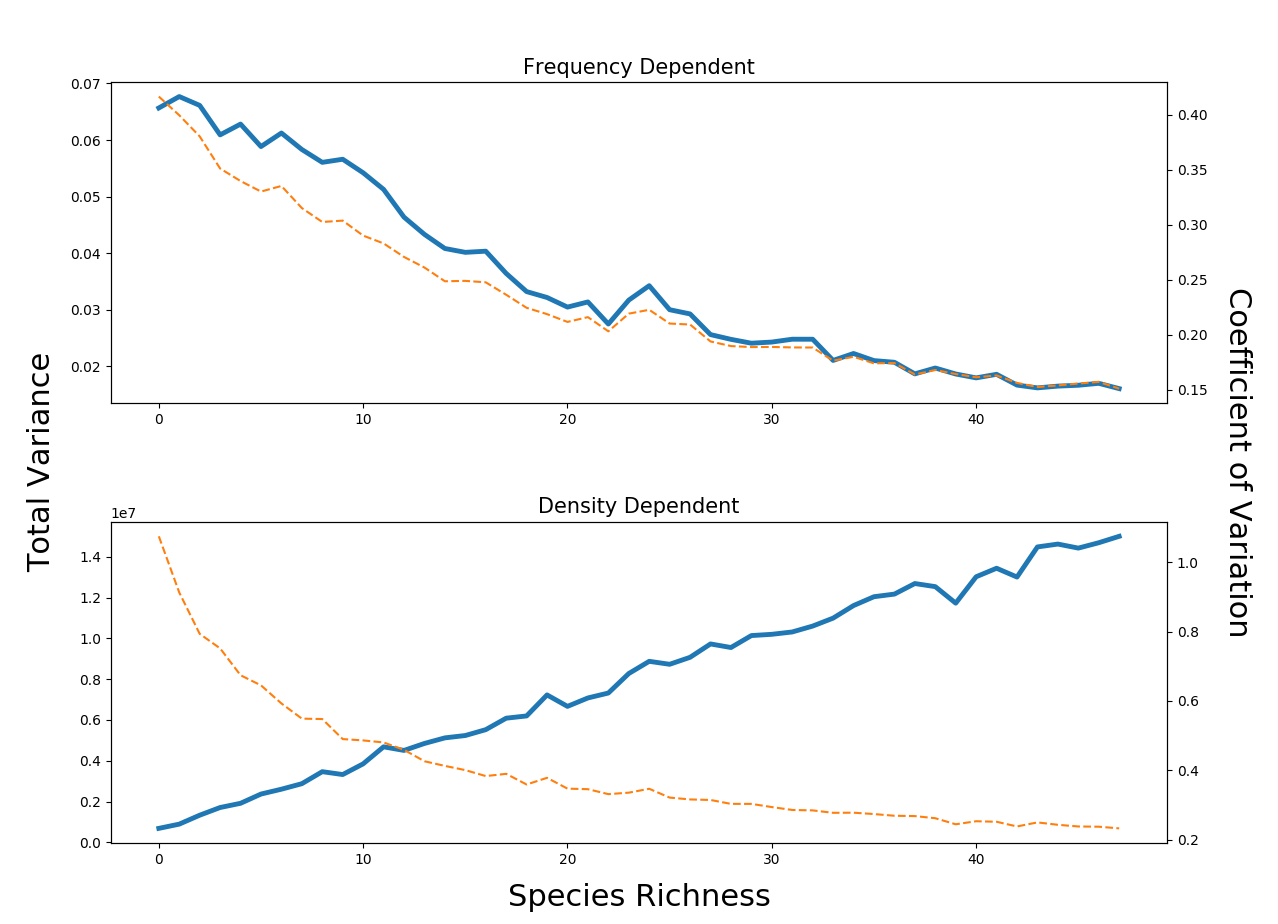}

  \caption{ Monte Carlo estimates of the total variance (blue, solid line) and the coefficient of variation (yellow, dashed line) of the community invasion factor, for frequency- and density-dependent transmission in the low $a_R$, low $c$ case.}
    \label{fig:rev1_plot}
\end{figure}

\rev{2}{3}{Finally, for the low $a_R$, low $c_{ij}$ case, and for each transmission mode, we consider two cases where the susceptibility of added host species increases or decreases with $n$, to better characterize how variance reduction behaves when community growth is not entirely random (Fig. \ref{fig:rev2_case}). In this case we set $a_R=0$, and instead sampled $R_i$ with from a truncated normal distribution (with support on $[0,2]$) with mean $E[R_i] \propto c_R n + 1$ where $c_R$ was set to $\pm .05$ for the positively and negatively correlated cases (respectively).} \rev{1}{3,4}{Fig \ref{fig:rev1_plot} shows an example of the behavior of the total variance, as well as the coefficient of variation for the low $c$, low $a_R$ case for both transmission modes.}

\begin{figure}
  \includegraphics[scale=.6]{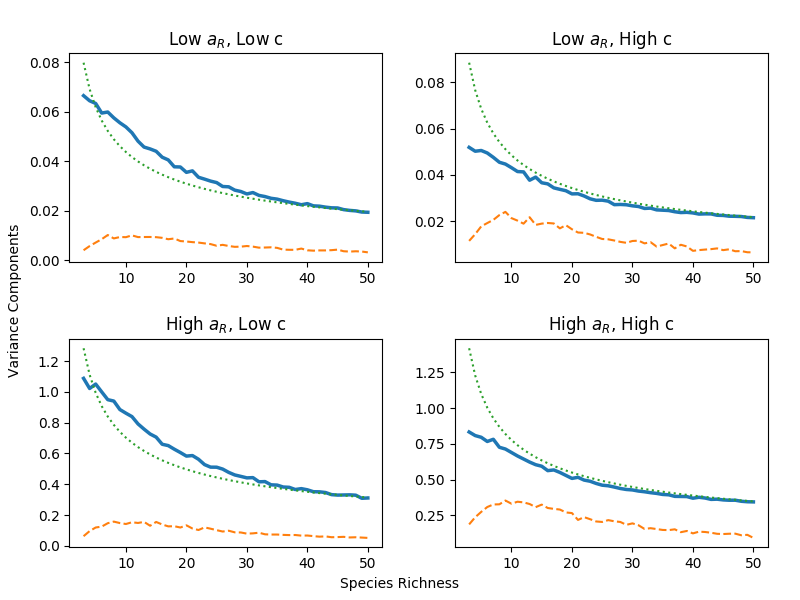}
  \caption{ Monte Carlo estimates of the mean conditional variance (blue, solid line) and the variance of the conditional mean (yellow, dashed line), the first and second terms (respectively) in Eq. (\ref{eq:totvar}), under the assumptions of the \textbf{frequency-dependent} NGM. In addition we plot the scaling behavior of the mean conditional variance predicted by its bound (green, dotted line). 
  We see that in all cases a variance reduction effect occurs in the mean conditional variance, as is predicted by Thm \ref{th:main}. Furthermore, in all cases we see that the mean conditional variance dominates the variance of the conditional mean, which explains why variance reduction is so pronounced in the sum of the two terms.}
  \label{fig:mc_freq}
\end{figure}

For the frequency-dependent NGM we see a clear variance reduction effect in every case (Fig. \ref{fig:mc_freq}). While the actual estimated values of the terms were substantially lower than the bounds given in Theorem 1 (which were not plotted for legibility), we do see that the predicted qualitative behavior holds.  Notably, in every case the mean conditional variance is strictly decreasing for $n>10$. Thus, under relatively low species diversity, one sees a variance amplification effect. Yet, in each case the variance of the conditional mean (which produced the constant term in Theorem 1) is smaller than the mean of the conditional variance. At higher diversity levels ($n > 10$), the sum of these two terms, the total variance of the community $R_0$, will therefore decrease rapidly in $n$: a variance reduction effect, as we can see in Fig \ref{fig:rev1_plot}.

\begin{figure}
  \includegraphics[scale=.6]{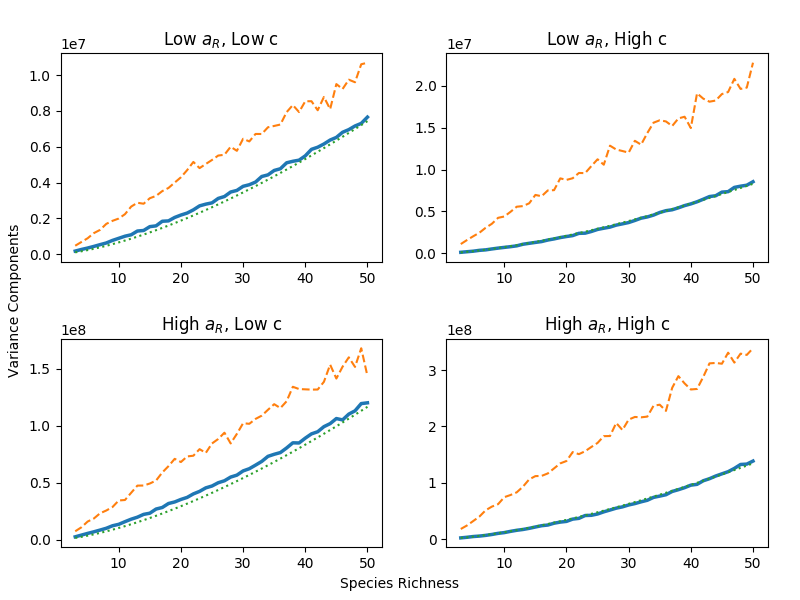}
  \caption{ Monte Carlo estimates of the mean conditional variance (blue, solid line) and the variance of the conditional mean (yellow, dashed line), the first and second terms (respectively) in Eq. (\ref{eq:totvar}), under the assumptions of the \textbf{density-dependent} NGM, as well as the predicted behavior of the mean conditional variance (green, dotted line, partially covered by the conditional variance in some plots). In this case variance reduction is overwhelmed by the $\mathcal{O}(n)$ scaling of the NGM due to density-dependence, and a variance amplification results.}
  \label{fig:mc_dense}
\end{figure}

As expected, we see that an opposite, variance amplification effect occurs for the density-dependent NGM (Fig. \ref{fig:mc_dense}).

This occurs because the effect of switching from frequency to density dependence is to multiply the NGM by a factor of $n$, which scales the variance by $n^2$. The concentration of measure is overwhelmed by this scaling, so the variance increases with $n$. \rev{2}{3}{We see a similar effect in Fig \ref{fig:rev2_case} when susceptibility and richness are positively correlated. Since the positive correlation case assumes that $R_i \sim \mathcal{O}(n)$, the reason for variance amplification is the same as for Fig. \ref{fig:mc_dense}. Note, however, that in the density dependent, negative correlation case we see that the variance amplification tapers for large $n$. This is because $R_i$ is bounded below by $0$, and so as $n$ grows that $R_i$ become small with high probability, counteracting the effect of density-dependence on the elements of the NGM.}

\rev{AE}{4}{Finally, in both Figs \ref{fig:mc_freq} and \ref{fig:mc_dense} we see that the qualitative scaling predicted by our bounds appears to hold for large $n$, even when the actual numerical value of our bound is much larger than the observed variance values. Importantly, however, in the frequency dependent case we see that the mean conditional variance decays slower than our bound would predict, and that for large $c$ the mis-fit becomes worse. This is predictable from the viewpoint of random matrix theory, as the off-diagonal terms of the NGM $G_n$ exhibit correlation with each other. Were all matrix elements independent, variance reduction of both the community invasion factor and the community $R_0$ would be guaranteed by the Wigner Semicircular law \citep{tao_random_2014}. The parameter $c$ has the effect of reducing the impact of these correlated parameters, so when $c$ is very small the matrix is more well-behaved, and the scaling behavior appears at smaller $n$.}

\section{Discussion}
\label{sec:discussion}
Our results have shown that, for frequency-dependent diseases, there exists terms $C_1 \sim \mathcal{O}(n^{-1/2})$ and $C_2 \sim \mathcal{O}(1)$ such that $\text{Var}[r(\pmb{R},\pmb{\theta},\pmb{K})] \leq C_1 + C_2$.  It is challenging to make claims about the relative size of $C_1$ and $C_2$ for a general randomized NGM, and in cases where $\pmb{R}$ is tightly determined by $\pmb{\theta}$ and $\pmb{K}$, $C_2$ may dominate the bound, so variance reduction may not appear.  However, in numerical experiments conducted with realistic distributions of $\pmb{R}$, $\pmb{\theta}$, and $\pmb{K}$ we observed that for values of $n$ less than $50$, $C_1$ dominates $C_2,$ and so variance reduction occurs. On the other hand, when disease is density-dependent our upper bound increases as $\mathcal{O}(n^{3/2})$ accordingly. In this case it is possible for either variance reduction or amplification to occur without violating our bound, although our numerical experiments suggests that variance amplification occurs for realistic biological parameter values.

\rev{2}{5}{
These results can be contrasted with the explanation of variance reduction found in \cite{mitchell_effects_2002}, which observed a variance reduction in their experimental data, and attributed it to decreasing variability in community composition, ie. a selection effect. Since the variance in community composition would presumably behave the same under either frequency or density dependent transmission, our results suggest that this explanation is incomplete. Rather, disease transmission mode mediates the relationship between community composition. In the case of frequency dependent transmission and high total community abundance, the introduction of a highly-competent host species will have a similar effect to the addition of a less competent host species, since the pathogen contact rate experienced by either will be fairly low. On the other hand, under density dependence the behavior of an epidemic will be more sensitive to community composition, and so a wider range of epidemic outcomes will be likely. The outcome will in part be determined by the probability of including one or more highly-competent hosts.

Mathematically this difference manifests in the presence or absence of an $\mathcal{O}(n^{-1})$ scaling factor in the NGM. The core of our theorem demonstrates that, in the density dependent case the matrix norm of our NGM concentrates in such a way that variance goes as $\mathcal{O}(n^{3/2)}$. Switching from density to frequency dependence then introduces the $\mathcal{O}(n^{-1})$, scaling the variance of the norm of the NGM such that it goes as $\mathcal{O}(n^{-1/2})$. This also suggests that any models where the NGM has elements which scale as $\mathcal{O}(n^{-1})$ (and where the elements are suitably independent) will lead to variance reduction. In Fig. \ref{fig:rev2_case} we see that scaling the $R_i$ with $n$ is sufficient to produce variance amplification, even when transmission is frequency-dependent, and that decreasing the $R_i$ with $n$ can counteract variance amplification when transmission is density-dependent. Furthermore, \cite{mihaljevic_scaling_2014} consider an identical model to ours but in addition distinguish between ``compensatory'' population growth, where $\sum_i K_i$ is held constant (and so $K_i \sim \mathcal{O}(n^{-1})$), and ``additive'' growth, where it is not. In numerical simulations they found that whenever population scaling was compensatory, variance reduction occurred, even if transmission was density dependent.}

There are a number of opportunities to improve or extend our results. A  simple improvement may be possible through a more careful calculation of the bound of $L_{\pmb{\theta},\pmb{K}}^2$ (see Appendix).  Other improvements are possible by considering a less general model of between-species infections. In the case of vector borne diseases, for example, a number of the $c_{ij}$ will go to zero, leaving only a handful of rows and columns nonzero which may allow for tighter control of the eigenvalues through the Gershgorin Circle Theorem. Alternately this could provide a route to derive a variance reduction effect for the community $R_0$. By setting enough $c_{ij}$ to zero such that the underlying host community is broken into non-interacting ``subcommunities'' that do not grow with $n$ to guarantee variance reduction in the spectral radius of $G_n$ \citep{schenker_semicircle_2005}. \rev{2}{7}{The results of \ref{subsubsec:spec_norm_rad} suggest that variance reduction may be expected for community $R_0$ in a more general setting purely by correlation with the community invasion factor (and similarly for metrics of epidemic severity such as pathogen prevalence).}

In addition to improving our bounds, they can be extended to cover several model variants. \rev{1}{1.1.1}{While the presented model assumes simple Malthusian population growth,  a more realistic extension would be to assume density-dependence in birth or death rates. For example,  changing $b_i$ to $b_i - \delta N_i$ in Eq. (\ref{eq:model})) would introduce density-dependence in birth rates. However, as long as the population size is at equilibrium when the epidemic begins, introducing density dependence into birth rate is not expected  to  affect our results, since this would only modifies $\dot{S}$, and the NGM at the core of our theorem depends only on $\dot{I}$. On the other hand, if we instead used density-dependent death rates (changing $d_i$ to $d_i + \delta N_i$) this 
    would also change the ``denominator" term of the NGM, resulting 
    in correlation between the $\theta_i$ and the $K_i$ factors. Depending on the severity of this correlation it may reduce or reverse the variance reduction, although the numerical experiments conducted in Figs. \ref{fig:mc_freq} suggest that variance reduction will occur even with some modest correlation between these parameters.} 

It is also possible to consider model variants where our bounds do not hold.  For example, if we reject our assumption that the triples $(R_i, \theta_i, K_i)$ are independent for different $i$.  While this assumption holds well for some experimentally designed host communities, it almost certainly does not hold for most natural ecosystems.  Ecological relationships between host species such as mutualism or predation can easily create correlations between the $K_i$, which may have implications for the relationships between the $\theta_i$ and $R_i$ as well.  While these correlations do not rule out a variance reduction effect necessarily, our bounds will not hold as proven and it would be difficult to make predictions about the ultimate behavior of the variance of community $R_0$.

While our model treats diversity as existing purely between homogeneous species, within-species heterogeneity due to genetic variation also can have substantial effects on epidemic dynamics \citep{elderd_developing_2013, flemingdavies_effects_2015, dwyer_host_1997}. Given that the community $R_0$ is still definable in these cases (indeed its original definition was explicitly for the case of a single, heterogeneous host species), we expect that phenomena similar to variance reduction can also occur due to genetic diversity, and that our results could be adapted to understand this case.

Using a dynamical system model of epidemic spread we have provided a mathematical framework for understanding the variance reduction effect in epidemics with multiple host species. Our results suggest that variance reduction is driven more by pathogen transmission mode (eg. density or frequency dependence), than by increased overlap between communities, ie. a selection effect. This work also shows the utility of random matrix theory to disease ecology more generally. Where the behavior of an epidemic in a diverse ecological community can be modeled by a dynamical system, we expect that randomized linearizations of that model can be a rich source of insight into disease-diversity phenomena.

\section*{Conflict of interest}
The authors declare that they have no conflict of interest.

\bibliography{ref.bib}
\bibliographystyle{spbasic}      

\newpage{}
\section{Appendix}
\label{sec:appendix}
\subsection{Bounding $L_{\theta,K}^2$}
\label{subsec:app_bound}
To bound $L_{\theta,K}^2$ we first observe that $\text{max}_{||R||_2 =1} r(R,\theta,K) = L_{\theta,K}$. We can therefore bound $L_{\theta,K}^2$ by bounding $r(R,\theta,K)^2$ for $R$ on the unit ball. Since $r(R,\theta,K)^2 \leq ||G'(R,\theta,K)||_{F}^2$, where $|| M ||_{F} = \sqrt{ \sum\limits_{i,j} M_{ij}^2}$ is the Frobenius matrix norm, we can compute an upper bound in terms of a sum of the matrix elements. Furthermore, we will use the fact that $||R||_1 \leq \sqrt{n} ||R||_2$, and define $\eta = \text{max}(\{1,\frac{c^2}{4}\}) * \frac{\Theta_2}{\Theta_1}$. This definition of $\eta$ is simple, but produces very loose bounds.  Improvements may be gained through a more careful consideration of this factor.

Now, from (\ref{eq:gprime}) and keeping in mind that $\sum_i R_i^2 = 1$, we have:
\[
\begin{split}
||G'(R,\theta,K)||_{F}^2 &= \sum_i \left( R_i^2 + \sum_j \frac{c_{ij}^2}{4}(R_i + \frac{\theta_j}{\theta_i} R_j)^2 \right)\\
&\leq \eta \left( \sum_i \left(  R_i^2 + \sum_j (R_i + \eta_1 R_j)^2 \right) \right)\\
&= \eta \left( \sum_i R_i^2 + \sum_{i,j \neq i} (R_i + R_j)^2 \right)\\
&= \eta \left( \sum_i R_i^2 + \sum_{i,j \neq i} (R_i^2 + 2 R_i R_j + R_j^2 ) \right)\\
&\leq \eta \left( \sum_i R_i^2 + \sum_{i,j \neq i} (R_i^2 + 2 R_j + R_j^2 ) \right)\\
&= \eta \left( \sum_i R_i^2 + (n-1) R_i^2 + 2( ||R||_1 - R_i ) + ||R||_2^2 -  R_i^2 ) \right)\\
&= \eta \left( ||R||_2^2 + (n-1) ||R||_2^2 + 2 n ||R||_1 - 2 ||R||_1 + n ||R||_2^2 -  ||R||_2^2 \right)\\
&\leq \eta \left( 1 + (n-1) + 2 n \sqrt{n} - 2 \sqrt{n} + n - 1 \right)\\
&= \eta \left( 2 n \sqrt{n} + 2n - 2 \sqrt{n} - 1  \right)\\
\end{split}
\]

Which proves the claim.

\subsection{Infection-Related Deaths}
In the simulations plotted in Fig. \ref{fig:rad_norm_size} we had assumed that no births or deaths occurred over the course of the epidemic. In particular, this excluded the possibility of infection-related deaths, a feature present in most models of multi-host infections. Here we assess the effect of this modeling choice on the results of our numerical simulations.

We first modify Eq (\ref{eq:model}) to include an infection related deaths term, $\sigma_i$:
Our multi-host SIR epidemic model is:
\begin{equation}
\begin{split}
\dot{S}_i &= b_i (S_i + I_i + D_i) - p_i \sum\limits_{j=1}^n \beta_{ij} I_j - d_i S_i\\
\dot{I}_i &= p_i \sum\limits_{j=1}^n \beta_{ij} I_j - (d_i + \alpha_i + \sigma_i) I_i\\
\dot{D}_i &= \alpha_i I_i - d_i D_i
\label{eq:inf_deaths_mod}
\end{split}
\end{equation}

\begin{figure}
    \centering
    \includegraphics[scale=.25]{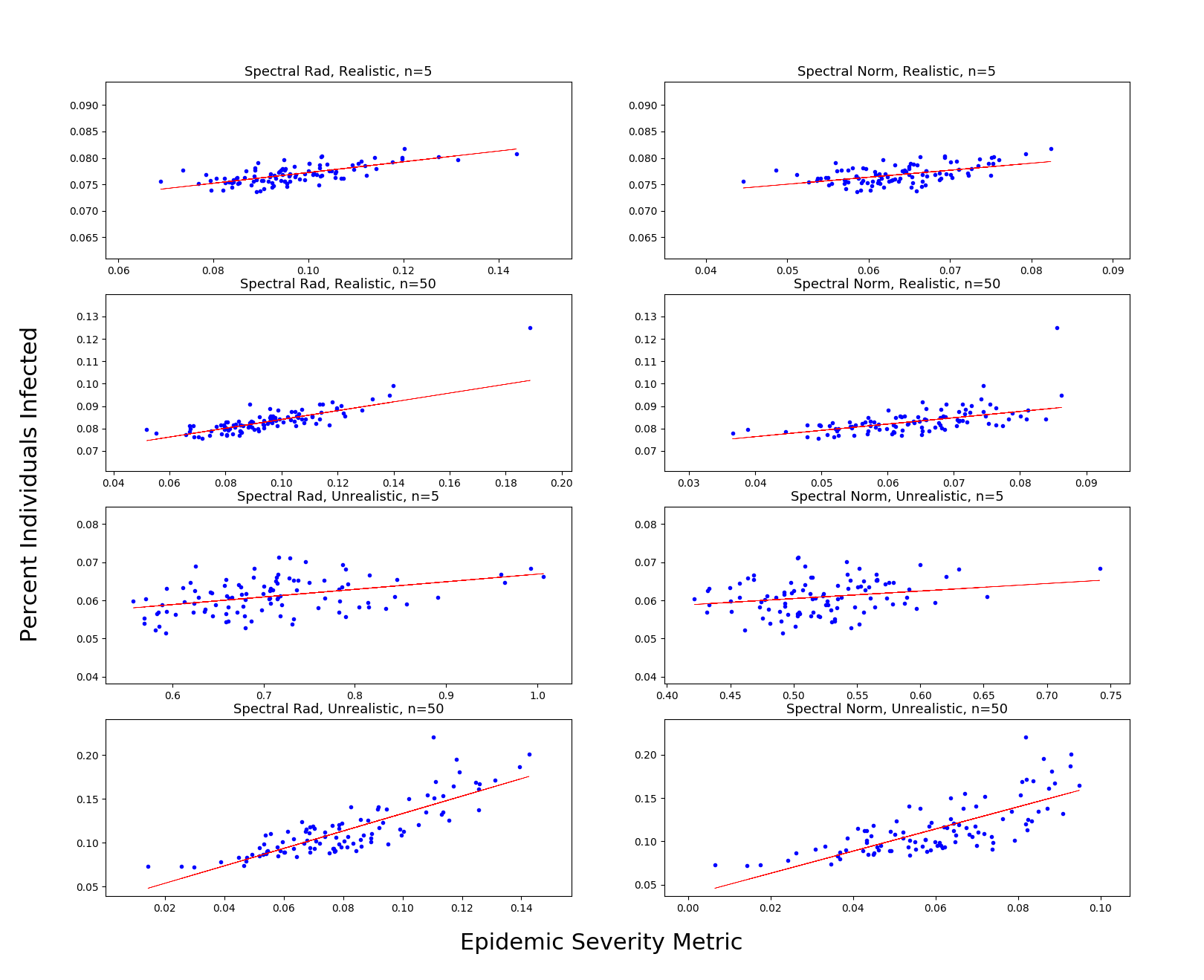}
    \caption{Relationships between epidemic size and $\rho(G_n)$ and $||G_n||$ with a model that includes infection-related deaths \ref{eq:inf_deaths_mod}. In all cases $c=.05$ and $a_R = .5$, and we set $\sigma_i = .5 d_i$ and $\alpha_i = 10 d_i$.}
    \label{fig:ae_fig}
\end{figure}

We assumed that $b_i = d_i$, that $\alpha_i = 10 d_i$, and that $\sigma_i = .5 d_i$. The results of this simulation are show in Fig \ref{fig:ae_fig}. We see that the primary effect of introducing $\sigma_i$ is to reduce average infection prevalence when $n=$, and to reduce the variability of prevalence in all cases. This has the effect of reducing the relationship between prevalence and both $\rho(G_n)$ and $||G_n$. Again we see that both perform similarly in predicting infection size, although $||G_n||$ still performs stightly worse, particularly when $n=50$.

\end{document}